\documentstyle[12pt,fleqn]{article}
\def\one{\mbox{{\sf 1}\kern-0.25em{\bf l}}}
\def\beq{\begin{equation}}
\def\eeq{\end{equation}}

\begin{document}

\newcommand{\KS}{Kochen-Specker }
\font\msbm=msbm10 scaled \magstep1 
\newcommand{\R}{\mbox{\msbm R}}  

\vspace*{15mm}
\begin{center}
{\Large Generalized Kochen-Specker Theorem$^*$} \vfill
Asher Peres\\[7mm]
{\sl Department of Physics, Technion -- Israel Institute of
Technology, 32 000 Haifa, Israel}\\[1cm]
\end{center}\vfill

\noindent{\bf Abstract}\bigskip

A generalized Kochen-Specker theorem is proved. It is shown that there
exist sets of $n$ projection operators, representing $n$ yes-no
questions about a quantum system, such that none of the $2^n$ possible
answers is compatible with sum rules imposed by quantum mechanics.
Namely, if a subset of commuting projection operators sums up to a
matrix having only even or only odd eigenvalues, the number of ``yes''
answers ought to be even or odd, respectively. This requirement may lead
to contradictions. An example is provided, involving nine projection
operators in a 4-dimensional space.

\vfill\noindent$^*$\,Dedicated to Professor Max Jammer, on the occasion
of his 80th birthday.\newpage

The \KS theorem [1] is of fundamental importance for quantum theory.
In its original form, it asserts that, in a Hilbert space with a finite
number of dimensions, $d\geq3$, it is possible to produce a set of $n$
projection operators, representing yes-no questions about a quantum
system, such that none of the $2^n$ possible answers is compatible with
the sum rules of quantum mechanics. Namely, if a subset of mutually
orthogonal projection operators sums up to the unit matrix, one and only
one of the corresponding answers ought to be yes. This requirement
cannot be fulfilled. The physical meaning
of this theorem is that there is no way of introducing noncontextual
``hidden'' variables [2] which would ascribe definite outcomes to these
$n$ yes-no tests. This conclusion holds irrespective of the preparation
(the quantum state) of the system being tested.

It is also possible to formulate a ``state-specific'' version of this
theorem, valid for systems that have been prepared in a known pure
state. In that case, the projection operators are chosen in a way
adapted to the known state. A smaller number of questions is then
sufficient to obtain incompatibility with the quantum mechanical sum
rules.  An even smaller number is needed if strict sum rules are
replaced by weaker probabilistic arguments [3, 4].

The original proof by Kochen and Specker [1] involved projection
operators over 117 vectors in a 3-dimensional real Hilbert space \R$^3$.
A simple proof with 33 vectors was later given by Peres [5], who also
reported an unpublished construction by Conway and Kochen, using only 31
vectors [6]. A proof with 20 vectors in \R$^4$ was given by Kernaghan
[7], and one with 30 projection operators in \R$^8$ by Kernaghan and
Peres [8].

There also is a multiplicative variant of the \KS theorem, first
mentioned by Fine and Teller [9]. Some explicit examples were given by
Mermin [10], after a particular state-specific case was discovered by
Peres [11]. The simplest one of theses examples involves nine
operators in four dimensions, whose eigenvalues are 1, 1, $-1$, and
$-1$. They can be written in terms of Pauli matrices for a pair of
spin-$1\over2$ particles, as follows:
\beq \begin{array}{rclcrclcrcl}
 \one & \!\!\!\otimes\!\!\! & \sigma_z & \quad & \sigma_z &
\!\!\!\otimes\!\!\!
& \one &  \quad & \sigma_z & \!\!\!\otimes\!\!\! & \sigma_z \\
  \sigma_x & \!\!\!\otimes\!\!\! & \one & \quad &  \one &
\!\!\!\otimes\!\!\! &
\sigma_x & \quad & \sigma_x & \!\!\!\otimes\!\!\! & \sigma_x
\rule{0pt}{2.7ex} \\
 \sigma_x & \!\!\!\otimes\!\!\! & \sigma_z & \quad & \sigma_z &
\!\!\!\otimes\!\!\! & \sigma_x &
\quad & \sigma_y & \!\!\!\otimes\!\!\! & \sigma_y \rule{0pt}{2.7ex}
\end{array} \eeq
Each one of these nine operators has eigenvalues $\pm 1$. In each row
and in each column, the three operators commute, and each operator is
the product of the two others, {\em except in the third column\/}, where
an extra minus sign is needed:

\beq (\sigma_x\otimes\sigma_x)\,(\sigma_y\otimes\sigma_y)
 =-(\sigma_z\otimes\sigma_z). \eeq
Because of that minus sign, it is impossible to attribute to the nine
elements of the above array numerical values, 1 or $-1$, which would be
the results of measurements of these operators (if such measurements
were performed), and which would obey the same multiplicative rule as
the operators themselves. We have reached a contradiction.

We thus see that what we call ``the result of the measurement of an
operator'' cannot in general depend only on the choice of that operator
and on the system being measured (unless that system is in an eigenstate
of that operator, or unless the operator itself is nondegenerate). The
ambiguous relationship between Hermitian operators and physical
observables complicates the epistemological meaning of the contradiction
that we have found. On the other hand, in the original (additive)
version of the \KS theorem, we only had to count the number of positive
answers in yes-no tests, and the physical meaning was clearer. In the
present article, I shall show how to convert a multiplicative \KS
contradiction into an ordinary, additive one.

It may seem that such a conversion is trivial: just take the logarithms
of the operators, and any product becomes a sum [9]. To find the
logarithm of an operator, we first transform it to a basis where that
operator is diagonal, we take the logarithms of its eigenvalues, and
then we transform back to the original basis. In particular, if the
eigenvalues are 1 and $-1$, they become 0 and $\pi i$, respectively.
Dividing the result by $\pi i$, we obtain a projection operator.
However, if we sum up the projection operators corresponding to any row
in array (1), we have a bad surprise: their product was the unit matrix,
but the sum of the logarithms does not vanish! In some cases it may be
$2\pi i$. With some hindsight, this could have been expected: a
logarithm is a multiply valued function, defined only modulo $2\pi i$.

There is however a way of overcoming this hurdle: instead of requiring
subsets of commuting projection operators to sum up to the unit matrix,
we merely require their sum to have all its eigenvalues with the {\it
same parity\/} (all even, or all odd). Since a ``measurement'' of this
sum, if performed, must yield one of the eigenvalues, this means, in
terms of our set of yes-no questions, that the number of positive
answers must have a definite parity. This may again lead to
contradictions.

Let us illustrate this with the nine operators in the above  array.
Their eigenvalues are $\pm1$, and each one of them can be converted into
a projection operator by the transformation ${\mit\Omega}\to(1-{\mit
\Omega})/2$. This is equivalent to taking the logarithm and dividing by
$\pi i$. For example, the first column of array (1) gives the three
projection operators:

\beq P_{2z}=(1-\sigma_{2z})/2, \eeq
\beq P_{1x}=(1-\sigma_{1x})/2, \eeq
\beq P_{1x2z}=(1-\sigma_{1x}\sigma_{2z})/2, \eeq
where 1 now means the 4-dimensional unit matrix, and we have discarded
the symbols $\one\,\otimes$ and $\otimes\,\one$, for brevity. Similar
abbreviated notations will also be used in the sequel.

We obtain

\beq P_{2z}+P_{1x}+P_{1x2z}=2-(1+\sigma_{2z})\,(1+\sigma_{1x})/2. \eeq
Since all these operators commute, and each parenthesis on the right
hand side of (\theequation) has eigenvalues 0 and 2, the eigenvalues of
the entire left hand side also are 0 and 2. This agrees with the fact
that the product of the three operators in the first column of our array
is equal to 1, so that its logarithm is 0 (mod $2\pi i$).

If we now attempt to attach to each one of these projection operators a
hypothetical numerical value, $v(P_{\ldots})=0$ or 1, we have

\beq v(P_{2z})+v(P_{1x})+v(P_{1x2z})=0\quad{\rm or}\quad2.
  \label{c1}\eeq
In the same way, we find, for the second column,

\beq v(P_{1z})+v(P_{2x})+v(P_{1z2x})=0\quad{\rm or}\quad2,
  \label{c2}\eeq
and for the first two rows,

\beq v(P_{2z})+v(P_{1z})+v(P_{1z2z})=0\quad{\rm or}\quad2,
  \label{r1}\eeq
and

\beq v(P_{1x})+v(P_{2x})+v(P_{1x2x})=0\quad{\rm or}\quad2.
  \label{r2}\eeq

For the third row, we have
\begin{eqnarray} P_{1x2z}+P_{1z2x}+P_{1y2y} & = &
 2-(1+\sigma_{1x}\,\sigma_{2z}+\sigma_{1z}\,\sigma_{2x}
  +\sigma_{1y}\,\sigma_{2y})/2, \\ & = &
 2-(1+\sigma_{1x}\,\sigma_{2z})\,(1+\sigma_{1z}\,\sigma_{2x})/2.
\end{eqnarray}
Again, the two parentheses on the right hand side commute, and each one
has eigenvalues 0 and 2. We thus have

\beq v(P_{1x2z})+v(P_{1z2x})+v(P_{1y2y})=0\quad{\rm or}\quad2.
  \label{r3}\eeq
Finally, for the third column of array (1), we have

\beq P_{1x2x}+P_{1y2y}+P_{1z2z}=
 2-(1+\sigma_{1x}\,\sigma_{2x}+\sigma_{1y}\,\sigma_{2y}
  +\sigma_{1z}\,\sigma_{2z})/2. \eeq
The eigenvalues of the rotationally invariant operator
$\sigma_{1x}\,\sigma_{2x}+\sigma_{1y}\,\sigma_{2y}
 +\sigma_{1z}\,\sigma_{2z}$ are well known: they are 1 for the triplet
state, and $-3$ for the singlet state. We thus have

\beq v(P_{1x2x})+v(P_{1y2y})+v(P_{1z2z})=1\quad{\rm or}\quad3.
  \label{c3}\eeq

The contradiction is now obvious: on the left hand sides of
Eqs.~(\ref{c1}--\ref{r2}), (\ref{r3}), and (\ref{c3}), each one of the
numbers $v(P_{\ldots})=0$ or 1 appears twice. The sum of these left hand
sides thus is an {\em even\/} number. On the other hand, the sum of
the right hand sides is odd. We thus obtain a \KS contradiction with
nine projection operators in a 4-dimensional space.

The above contradiction is an algebraic property of these nine
operators, irrespective of the quantum state of the physical system.
However, if it is known that the latter has been prepared in a
particular quantum state, for example the singlet state, a contradiction
may be obtained with fewer operators [11]. Consider those in the first
two columns of our array, so that Eqs.~(\ref{c1}) and (\ref{c2}) still
hold. For a singlet state, we also have

\beq (\sigma_{1j}+\sigma_{2j})\,\psi=0, \eeq
and therefore

\beq v(P_{1x})+v(P_{2x})=1, \label{sx} \eeq
\beq v(P_{1z})+v(P_{2z})=1. \label{sz} \eeq
Moreover, for a singlet

\beq (\sigma_{1x}\,\sigma_{2z}+\sigma_{1z}\,\sigma_{2x})\,\psi=0,
 \label{singl} \eeq
as may easily be verified by placing a factor

\beq 1\equiv
  (\sigma_{1x}\,\sigma_{2z})^{-1}\,(\sigma_{1x}\,\sigma_{2z}), \eeq
on the left of Eq.~(\ref{singl}). Therefore, the hypothetical values of
the corresponding projection operators must satisfy

\beq v(P_{1x2z})+v(P_{1z2x})=1,  \eeq
just as in Eqs.~(\ref{sx}) and (\ref{sz}).

Consider now Eqs.~(\ref{c1}), (\ref{c2}), (\ref{sx}), (\ref{sz}), and
(\theequation). On their left hand sides, each one of the numbers
$v(P_{\ldots})=0$ or 1 appears twice. The sum of these left hand sides
is an {\em even\/} number, just as before, while the sum of their right
hand sides is odd. This is a \KS contradiction involving only six
projection operators in a 4-dimensional space. It is however restricted
to a particular singlet state.

\bigskip I am grateful to N. D. Mermin for patiently explaining to me
that ref.~[11] was a \KS argument, not one about locality, as I had
wrongly thought. This work was supported by the Gerard Swope Fund, and
the Fund for Encouragement of Research.\clearpage

\frenchspacing
\begin{enumerate}
\item S. Kochen and E. P. Specker, J. Math. Mech. 17 (1967) 59.
\item M. Redhead, Incompleteness, nonlocality, and realism (Clarendon
Press, Oxford, 1987).
\item R. Clifton, Am. J. Phys. 61 (1993) 443.
\item H. B. Johansen, Am. J. Phys. 62 (1994) 471.
\item A. Peres, J. Phys. A: Math. Gen. 24 (1991) L175.
\item A. Peres, Quantum theory: concepts and methods (Kluwer, Dordrecht,
1993) p. 114.
\item M. Kernaghan, J. Phys. A: Math. Gen. 27 (1994) L829.
\item M. Kernaghan and A. Peres, Phys. Letters A 198 (1995) 1.
\item A. Fine and P. Teller, Found. Phys. 8 (1978) 629.
\item N. D. Mermin, Phys. Rev. Lett. 65 (1990) 3373; Rev. Mod. Phys. 65
(1993) 803.
\item A. Peres, Phys. Letters A 151 (1990) 107; Found. Phys. 22 (1992)
357.
\end{enumerate}

\end{document}